# Integrating Physics-Informed Neural Networks and 3D Vascular Geometry Learning for Cerebral Aneurysm Detection and Multimodal Rupture-Risk Prediction

Eshan Vipuil[1] and Xianqi Li[2,*]

[1]*West Shore Jr/Sr High School, Melbourne, FL, 32935, USA*

[2]*Department of Mathematics and Systems Engineering, Florida Institute of Technology, Melbourne, FL, 32901, USA*

[*]Corresponding author: xli@fit.edu

## Abstract

Cerebral aneurysms are localized dilations of intracranial arteries that may rupture and cause subarachnoid hemorrhage. Current assessment relies on human interpretation of imaging and clinical risk factors, but integrating vascular shape, flow-related information, and patient-level variables into a unified quantitative model remains challenging. This study develops a modular framework for cerebral aneurysm detection and rupture-risk prediction using 3D vascular geometry learning, physics-informed hemodynamic descriptors, and clinical variables. A PointNeXt-based detector first identified aneurysm presence from vascular point clouds. For aneurysm-positive cases, an unsteady physics-informed neural network then generated geometry-conditioned pressure, velocity, wall shear stress (WSS), time averaged WSS, oscillatory shear index (OSI), and relative residence time descriptors under prescribed Navier-Stokes residual and boundary-condition constraints. Multimodal models then integrated vascular morphology, physics-informed hemodynamic descriptors, and clinical variables to produce rupture-risk scores. The aneurysm detector achieved pooled out-of-fold area under the receiver operating characteristic curve (AUROC) of 0.959 and area under the precision-recall curve (AUPRC) of 0.859. For rupture-risk prediction, fixed 70/30 late fusion achieved the highest performance among evaluated models, with pooled AUROC of 0.827 and AUPRC of 0.732, exceeding all comparison models after Holm-corrected paired DeLong testing (all adjusted $p < 0.05$). Feature analysis identified OSI distribution, aneurysm location, radial geometry, and TAWSS descriptors as important contributors to cross-sectional rupture-risk discrimination. Together, these results provide a quantitative, multimodal strategy for case-specific aneurysm assessment.

## 1. Introduction

Cerebral aneurysms are localized dilations of intracranial blood vessels that may rupture and cause subarachnoid hemorrhage, a life-threatening event associated with substantial morbidity and mortality (Mayo Clinic, 2024). Many unruptured aneurysms remain asymptomatic and may never rupture during a patient's lifetime, creating a difficult clinical management problem. Preventive treatment through endovascular intervention or microsurgical clipping can reduce future hemorrhage risk in selected patients, but these procedures also carry non-negligible procedural risks (Brown & Broderick, 2014; Thompson et al., 2015). Current assessment relies on radiographic imaging modalities such as computed tomography angiography (CTA), magnetic resonance angiography (MRA), and digital subtraction angiography (DSA), together with clinician interpretation of aneurysm morphology and patient-level factors, including age, biological sex, aneurysm location, and prior history (Greving et al., 2014; Thompson et al., 2015). Although these workflows provide essential clinical information, aneurysm rupture reflects coupled effects of vascular morphology, local hemodynamics, vessel-wall properties, and patient-specific physiology. This complexity motivates quantitative approaches that can integrate complementary information sources for case-specific aneurysm assessment.

Deep learning (DL) has emerged as a promising data-driven approach for analyzing cerebral aneurysms from medical imaging and vascular geometry. Recent work spans DL-assisted aneurysm detection and 3D dataset development (Park et al., 2019; Yang et al., 2020), integrated CTA detection, segmentation, and morphometric analysis (Yang et al., 2025); CTA, radiomics, and multi-omics stability or rupture-status models (Feng et al., 2023; Li et al., 2023; Zeng et al., 2025; Shi et al., 2026); point-cloud rupture-status models (Luo et al., 2023; Cao et al., 2024); and clinical or morphology-based machine learning (Johnson et al., 2025; Feng et al., 2025; Kofman et al., 2026). Recent systematic reviews conclude that performance is promising but that most studies remain retrospective, heterogeneous, and insufficiently calibrated or externally validated (Daga et al., 2025; Maroufi et al., 2025; Owens et al., 2025; Yan et al., 2026). However, many existing DL models primarily learn rupture-associated imaging, radiomic, or geometric patterns, while explicit flow-related information is either omitted or incorporated through computationally expensive simulation-derived features. This motivates learning frameworks that can integrate vascular geometry, clinical variables, and hemodynamic descriptors in a unified rupture-risk model.

Computational fluid dynamics (CFD) has played a central role in aneurysm biomechanics by enabling patient-specific simulation of pressure, velocity, wall shear stress (WSS), time averaged WSS (TAWSS), oscillatory shear index (OSI), and relative residence time (RRT) (Cebral et al., 2005; Dhar et al., 2008; Shojima et al., 2004; Sforza et al., 2009). These descriptors provide complementary information about local flow behavior that vascular geometry alone does not directly encode. However, recent reviews continue to emphasize conflicting rupture associations, sensitivity to boundary conditions, meshing and solver choices, and barriers to clinical translation (Loly et al., 2025; Goudarzian et al., 2025; Valen-Sendstad & Steinman, 2014). Reduced-order and geometry-aware learning methods can accelerate CFD-derived field prediction, but current examples depend on simulation labels and show variable generalization across quantities, particularly OSI (Rajhi et al., 2025; Sheng et al., 2026). Physics-informed neural networks (PINNs) provide a different strategy by incorporating governing equations into the training objective and constraining neural network predictions through fluid-mechanics residuals and boundary-condition losses (Karniadakis et al., 2021; Raissi et al., 2019). In cerebral aneurysm modeling, this permits geometry-conditioned pressure-, velocity-, and WSS-derived descriptors without paired CFD field supervision.

These considerations motivate a multimodal framework that combines vascular shape, flow-related descriptors, and clinical variables. In this study, multimodal learning refers to the integration of three complementary information streams: vascular geometry encoded from 3D point clouds, PINN-derived hemodynamic descriptors generated under physical constraints, and clinical variables derived from patient- and aneurysm-level metadata. Geometry describes aneurysm and parent-vessel morphology, hemodynamic descriptors summarize local flow behavior, and clinical variables provide patient-level and anatomical context. Because the present task uses observed rupture status as a cross-sectional endpoint, the resulting risk scores should be interpreted as rupture-risk discrimination rather than direct estimates of future rupture timing or individualized prospective rupture probability. Interpretability is therefore important for identifying which morphology-, flow-, and clinical-related descriptors contribute to the model output while avoiding overinterpretation of any single hemodynamic mechanism. This distinction is important because published studies do not support a universal WSS, OSI, or RRT pattern across all aneurysm locations and populations (Lv et al., 2021; Xiang et al., 2011).

To address these gaps, this study makes the following contributions:

- Develops a staged detection-to-risk framework that links 3D point-cloud aneurysm screening with multimodal rupture-risk prediction.
- Introduces PINN-derived pressure, velocity, WSS, TAWSS, OSI, and RRT as physics-constrained hemodynamic descriptors without paired CFD supervision.
- Integrates vascular geometry, hemodynamic descriptors, and clinical variables to produce case-level rupture-risk scores.
- Quantifies the added value of each modality through matched comparisons of geometry-only, clinical-only, flow-augmented, early-fusion, and late-fusion models.
- Assesses late-fusion robustness and identifies key morphology-, flow-, and clinical-related contributors to cross-sectional rupture-risk discrimination.

We contrast the current clinician-led pathway with the proposed conditional detection-to-rupture-risk workflow in **Figure 1**.

The remainder of this paper is organized as follows. Section 2 describes the study cohorts, vascular geometry processing, PointNeXt-based aneurysm detection, physics-informed hemodynamic descriptor generation, and multimodal rupture-risk prediction models. Section 3 presents the experimental setting, including training protocols, evaluation metrics, statistical testing, and feature-importance analysis. Section 4 reports the results for aneurysm detection, PINN optimization behavior, geometry-only and multimodal rupture-risk prediction, and descriptor-level feature analysis. Section 5 discusses the main findings, limitations, and future directions. Section 6 concludes the paper.

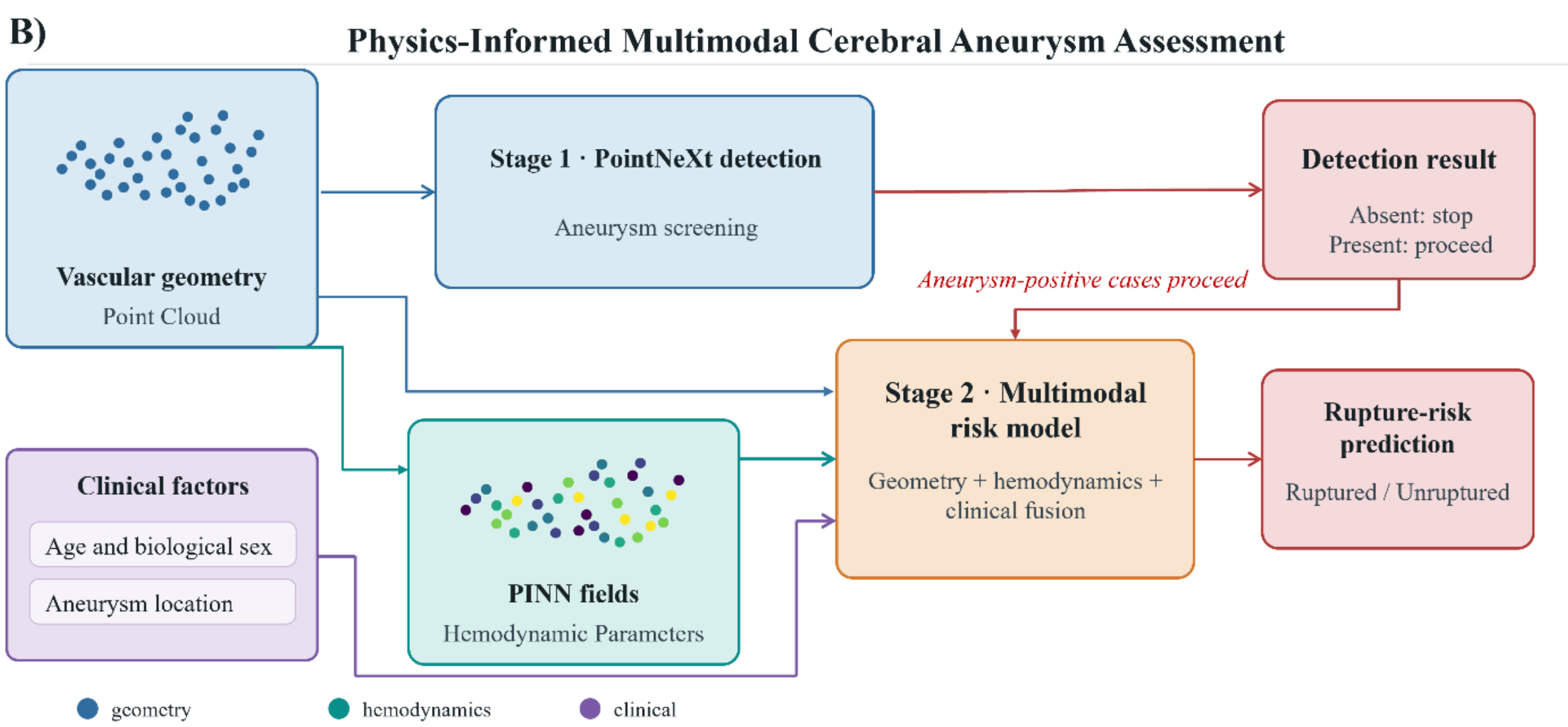

**Figure 1. Conventional and proposed cerebral aneurysm assessment workflows**. **A**) The current clinician-led pathway integrates angiographic imaging and clinical factors for aneurysm detection, rupture assessment, and management decisions. **B**) The proposed deep learning pathway first uses PointNeXt to detect aneurysm presence. Aneurysm-absent cases stop after screening, whereas aneurysm-positive cases proceed to multimodal rupture-risk prediction using geometry, PINN-derived hemodynamic descriptors, and clinical factors. The illustrative digital subtraction angiogram in A) is adapted from Lucien Monfils (2000), Wikimedia Commons, CC BY-SA 3.0.

## 2. Methods

### *2.1 Study Cohorts and Prediction Tasks*

This study evaluated two related but independently trained prediction tasks: 1) Aneurysm detection from vascular surface geometry; 2) Rupture-risk prediction among aneurysm cases. Each task used a separate open vascular-geometry cohort. For aneurysm detection, surface models were obtained from the Vascular Model Repository (VMR), which provides normal and diseased cardiovascular geometries in formats compatible with the SimVascular workflow (Schroeder et al., 2006; Updegrove et al., 2017). The detection cohort included 304 VTK PolyData (.vtp) geometries, comprising 75 aneurysm-positive geometries and 229 non-aneurysm geometries. The non-aneurysm group included normal vascular models and other vascular disease geometries.

For rupture-risk prediction, aneurysm geometries and metadata were obtained from AneuX, which consolidates cases from the AneuX, @neurIST, and Aneurisk projects and provides 750 aneurysm domes linked to 668 vessel trees at multiple surface resolutions and vessel-cut configurations (Juchler et al., 2022). The selected vessel-cut geometry retained the aneurysm sac and attached parent-vessel segment. Available AneuX metadata included rupture status, aneurysm location, side, age, and biological sex. Of the 750 modeled aneurysms, 735 had documented rupture status and formed the common rupture-status cohort, including 261 ruptured and 474 unruptured aneurysms. 15 cases with blank or unknown rupture status were excluded from all rupture-risk metrics, statistical testing, and feature analyses.

Observed rupture status was used as a cross-sectional proxy endpoint for rupture-risk discrimination. Accordingly, the models produced case-level rupture-risk scores associated with observed ruptured-versus-unruptured status. Rupture-risk models used vascular geometry, PINN-derived hemodynamic descriptors, and available clinical variables, including age, biological sex, aneurysm location, and side. These inputs were selected to reflect the established role of patient-level factors, aneurysm location, vascular morphology, and hemodynamic environment in aneurysm assessment (Brown & Broderick, 2014; Etminan et al., 2015; Greving et al., 2014; Morita et al., 2012; Thompson et al., 2015; Wiebers et al., 2003). All rupture-risk models were evaluated on the same 735 labeled aneurysm cases to enable paired comparison across geometry-only, clinical-only, flow-augmented, and multimodal modeling strategies.

### *2.2 Vascular Geometry Learning and Aneurysm Detection*

All vascular geometries were provided as VTK PolyData (.vtp) surface models and converted into point-cloud representations by sampling xyz coordinates from the vascular mesh. Point clouds were used because they allow direct modeling of irregular three-dimensional vascular surfaces without voxelization or mesh remeshing, while preserving local and global vascular shape information. For aneurysm detection, each Vascular Model Repository geometry was sampled to a fixed-size point cloud of 2,048 surface points and normalized before being passed to a PointNeXt-based detector. The detector classified each geometry as aneurysm-positive or non-aneurysm. PointNeXt was selected because it extends PointNet and PointNet++ with improved hierarchical point-cloud feature extraction and training strategies, making it suitable for learning local-to-global vascular morphology from irregular surface data (Qi et al., 2017a, 2017b; Qian et al., 2022). Random-start farthest-point sampling was used during training to increase sampling variability, whereas deterministic sampling was used during evaluation to ensure reproducible predictions.

The detector used two set-abstraction stages to progressively aggregate local vascular features. The first stage selected 512 centroids with a neighborhood radius of 0.12, and the second stage selected 128 centroids with a neighborhood radius of 0.24. Within each stage, local neighborhoods were formed by radius- or k-nearest-neighbor grouping. Relative xyz coordinates were combined with incoming point features and encoded using shared pointwise multilayer perceptrons with Gaussian error linear unit activations. Symmetric maximum pooling produced order-invariant local descriptors, followed by global pooling to generate a 1,024-dimensional case-level representation. This representation was passed through a fully connected classifier with 512 and 256 hidden units to produce aneurysm-positive versus non-aneurysm logits. The corresponding detection network architecture is illustrated in Supplementary Information **Figure S1.**

For rupture-risk prediction, the selected AneuX cut geometries were processed using a consistent point-cloud representation. Surface coordinates were centered at the case centroid and scaled by the maximum radial distance to fit within a unit sphere. These normalized point clouds were used as geometric inputs for the rupture-risk prediction models. The same processed AneuX vascular surfaces also defined the geometric domains for PINN-based hemodynamic descriptor generation. Interior collocation points, wall points, inlet points, and outlet points were sampled from each processed geometry as described in Section 2.3. Thus, vascular geometry provided the structural basis for both PointNeXt-based morphology learning and PINN-derived hemodynamic descriptor extraction in the rupture-risk modeling stage.

### *2.3 PINN Hemodynamic Field Generation and Descriptor Extraction*

To provide flow-related information for rupture-risk prediction, PINNs were used to generate geometry-conditioned hemodynamic descriptors from each aneurysm surface under prescribed modeling assumptions. PINNs constrain neural-network predictions through the residuals of governing partial differential equations and have been widely used in scientific machine learning and fluid-mechanics applications (Cuomo et al., 2022; Karniadakis et al., 2021; Raissi et al., 2019). Each aneurysm geometry was processed to define the vascular domain, wall surface, inlet boundary, and outlet boundary. Following established cardiovascular modeling workflows (Cebral et al., 2005; Taylor & Figueroa, 2009; Updegrove et al., 2017), the inlet was selected as the largest open vessel cap, and the inflow direction was estimated from the local cap normal. No-slip velocity was imposed on the vessel wall, and zero relative pressure was imposed at the outlet to define the pressure gauge. Interior collocation points, wall points, inlet points, and outlet points were sampled from each processed geometry to evaluate the physics and boundary-condition losses.

For each case, the PINN predicted dimensionless relative pressure $p$ and 3D velocity $\boldsymbol{u}$, which are stacked as $\boldsymbol{q}(\boldsymbol{r},t) = [p(\boldsymbol{r},t), \boldsymbol{u}(\boldsymbol{r},t)]$, $\boldsymbol{u}(\boldsymbol{r},t) = [u_\mathrm{x}(\boldsymbol{r},t), u_\mathrm{y}(\boldsymbol{r},t), u_\mathrm{z}(\boldsymbol{r},t)]$, where $\mathbf{r} = (\mathrm{x},\mathrm{y},\mathrm{z})$ denotes spatial position and $t$ denotes time within one cardiac cycle. The continuity and momentum constraints were treated together as the incompressible normalized Navier-Stokes system:

$$\begin{gathered} \nabla \cdot \boldsymbol{u} = 0 \\ \frac{\partial \boldsymbol{u}}{\partial t} + (\boldsymbol{u} \cdot \nabla)\boldsymbol{u} + \nabla p - \kappa \nabla^2 \boldsymbol{u} = \boldsymbol{0} \end{gathered} \tag{1}$$

where $\kappa = \frac{1}{\Re}$ is the inverse Reynolds number used in the normalized momentum residual. The Reynolds number was defined as $\Re = \frac{\rho U L}{\mu}$, where $\rho$ is blood density, $U$ and $L$ are case-specific velocity and length scales, and $\mu$ is dynamic viscosity. A prescribed pulsatile inlet waveform was used, $s(t) = 1 + 0.6\sin\left(\frac{2\pi t}{T}\right)$, where $T$ denotes the cardiac-cycle duration and the coefficient 0.6 controls the waveform amplitude about its unit mean. The PINN was optimized using five loss components: the Navier–Stokes physics residual, inlet velocity loss, wall no-slip loss, outlet pressure loss, and periodicity loss. The residual loss was defined as

$$\mathcal{L}_\mathrm{phys} = \langle \mathcal{R}_u^2 \rangle + \langle \mathcal{R}_v^2 \rangle + \langle \mathcal{R}_w^2 \rangle + \langle \mathcal{R}_\mathrm{cont}^2 \rangle, \tag{2}$$

where $\mathcal{R}_u$, $\mathcal{R}_v$, and $\mathcal{R}_w$ are the momentum-residual components and $\mathcal{R}_\mathrm{cont}$ is the incompressibility residual. The boundary and periodic losses were grouped together:

$$\begin{gathered} \mathcal{L}_\mathrm{inlet} = \langle \|\boldsymbol{u}(\boldsymbol{r},t) - \boldsymbol{u}_\mathrm{inlet}(t)\|^2 \rangle_{x \in \Gamma_\mathrm{inlet}}, \\ \mathcal{L}_\mathrm{wall} = \langle \|\boldsymbol{u}(\boldsymbol{r},t)\|^2 \rangle_{x \in \Gamma_\mathrm{wall}}, \\ \mathcal{L}_\mathrm{outlet} = \langle |p(\boldsymbol{r},t) - p_\mathrm{out}|^2 \rangle_{x \in \Gamma_\mathrm{outlet}}, \\ \mathcal{L}_\mathrm{periodic} = \langle \|\boldsymbol{q}(\boldsymbol{r},0) - \boldsymbol{q}(\boldsymbol{r},T)\|^2 \rangle_{x \in \Omega}, \end{gathered} \tag{3}$$

where $\Gamma_\mathrm{inlet}$, $\Gamma_\mathrm{wall}$, and $\Gamma_\mathrm{outlet}$denote the inlet, wall, and outlet boundaries, respectively; $\Omega$ denotes the vascular domain; Angle brackets $\langle \cdot \rangle$denote means over sampled points. The relative outlet pressure was set to $p_\mathrm{out} = 0$ to define the relative pressure gauge. This condition fixes the arbitrary pressure offset in the incompressible formulation and should be interpreted as a reference pressure, not as an assumption of zero absolute physiological pressure.

During optimization, each loss component was normalized by its initial reference magnitude to reduce scale imbalance among constraints, as shown:

$$\begin{gathered} \mathcal{L}_\mathrm{total} = \lambda_\mathrm{phys} \frac{\mathcal{L}_\mathrm{phys}}{\mathcal{L}_\mathrm{phys}^0} + \lambda_\mathrm{inlet} \frac{\mathcal{L}_\mathrm{inlet}}{\mathcal{L}_\mathrm{inlet}^0} + \lambda_\mathrm{wall} \frac{\mathcal{L}_\mathrm{wall}}{\mathcal{L}_\mathrm{wall}^0} + \lambda_\mathrm{outlet} \frac{\mathcal{L}_\mathrm{outlet}}{\mathcal{L}_\mathrm{outlet}^0} + \\ \lambda_\mathrm{periodic} \frac{\mathcal{L}_\mathrm{periodic}}{\mathcal{L}_\mathrm{periodic}^0}. \end{gathered} \tag{4}$$

The *lambda* terms are constraint weights, and each reference magnitude is an exponential moving average of the corresponding loss component. The reported hemodynamic fields used fixed weights of 1 for physics, 5 for inlet, 5 for wall, 2 for outlet, and 1 for periodicity. Inlet and wall constraints received greater emphasis because they define the imposed flow and no-slip boundary behavior, whereas the outlet term anchors the relative pressure gauge. These values were selected empirically during manual pilot tuning after component normalization. Automatic learning of these weight parameters will be investigated in future work. A schematic of the PINN architecture is provided in **Figure S2**. Training settings and unit-conversion procedures are described in Section 3.1.

After training, the retained outputs included relative pressure, instantaneous and time-averaged velocity, instantaneous wall shear stress, time-averaged TAWSS, OSI, relative residence RRT. Velocity magnitude was computed as $\|\boldsymbol{u}\| = \sqrt{u_x^2 + u_y^2 + u_z^2}$, The wall-shear vector was approximated from the near-wall tangential velocity gradient, $\boldsymbol{\tau}_w = \mu \frac{\partial \boldsymbol{u}_t}{\partial n}$, where $\boldsymbol{u}_t$ is the tangential velocity, $n$ is the outward wall-normal direction, and $\mu$ is dynamic viscosity. The wall-shear magnitude and TAWSS were computed as $\mathrm{TAWSS} = \frac{1}{T}\int_0^T |\boldsymbol{\tau}_w(t)|\, dt$ with $|\boldsymbol{\tau}_w| = \|\boldsymbol{\tau}_w\|$. The oscillatory shear index was defined as

$$\mathrm{OSI} = \frac{1}{2}\left(1 - \frac{\left\|\int_0^T \boldsymbol{\tau}_w(t)dt\right\|}{\int_0^T |\boldsymbol{\tau}_w(t)|\, dt}\right) \tag{5}$$

and relative residence time was computed as

$$\mathrm{RRT} = \frac{1}{(1-2\mathrm{OSI})\mathrm{TAWSS}}. \tag{6}$$

The retained PINN-derived descriptors were used by downstream rupture-risk models either as per-point flow channels paired with the vascular point cloud or as engineered case-level summary features for feature-importance analysis.

### *2.4 Rupture-Risk Prediction Models*

Rupture-risk prediction was formulated as a supervised binary prediction task among aneurysm cases with documented rupture status. Each model produced a case-level probability for the ruptured class, interpreted as a rupture-risk score. All models were evaluated on the same 735 AneuX cases, allowing paired comparison across different input modalities and fusion strategies. The model comparison was designed to assess whether vascular geometry alone contains rupture-risk information, whether clinical variables and PINN-derived hemodynamic descriptors provide complementary information, and whether early or late multimodal fusion improves predictive discrimination. For convenience, a workflow for this section is provided in **Figure S3.**

#### *2.4.1 Geometry-Only and Clinical Baseline Models*

Geometry-only models were first evaluated to establish the discriminatory value of vascular morphology alone. Four three-dimensional representations were compared: PointNet++, PointNeXt, a voxel-based 3D convolutional neural network, and a graph neural network (GNN) (Çiçek et al., 2016; LeCun et al., 2015; Qi et al., 2017b; Qian et al., 2022; Veličković et al., 2018;

Wang et al., 2019). All geometry-only inputs used the selected AneuX cut surface, which retained the aneurysm sac and attached parent-vessel segment. Point-cloud coordinates were centered at the case centroid and scaled by the maximum radial distance so that each geometry fit within a unit sphere.

The point-cloud baselines directly processed sampled xyz surface points. PointNet++ used hierarchical set-abstraction layers to group local vascular neighborhoods and aggregate local geometric features into a global case-level descriptor. PointNeXt used residual point-cloud blocks with local neighborhood grouping and global pooling to generate a compact geometry embedding. PointNeXt was used as the primary geometry encoder for multimodal prediction because it preserves local-to-global vascular shape information while providing a scalable point-cloud architecture for irregular surface data. The voxel CNN converted each normalized surface into a binary occupancy grid and learned volumetric shape features using residual 3D convolutional blocks. The GNN represented each aneurysm surface as a k-nearest-neighbor graph constructed from sampled surface points and used graph-attention layers to aggregate local structural relationships. The graph was built from point-neighbor relations rather than original VTP mesh connectivity to provide a consistent graph representation across cases.

A clinical-only model was included to quantify the predictive information contained in patient- and aneurysm-level metadata. The clinical encoder used age, biological sex, aneurysm location, and side. Age was treated as a continuous variable and standardized. Biological sex was numerically encoded, while aneurysm location and side were one-hot encoded. Missing ages were replaced with the cohort median, and missing side labels were represented by an explicit Unknown category. The encoded clinical vector was processed by a multilayer perceptron to produce a clinical embedding and a case-level rupture-risk probability.

### *2.4.2 Flow-Augmented and Multimodal Fusion Models*

Multimodal models were constructed to evaluate whether PINN-derived hemodynamic descriptors and clinical variables provided information beyond vascular geometry. The Flow + Geometry model augmented the PointNeXt geometry encoder with PINN-derived per-point hemodynamic channels. For each sampled surface point, normalized coordinates were paired with flow-related descriptors based on WSS, TAWSS, OSI, RRT, and their threshold or interaction terms. This model tested whether physics-constrained hemodynamic information improved rupture-risk prediction beyond morphology alone.

The Geometry + Clinical model combined the PointNeXt geometry embedding with the clinical embedding before classification. This model was included to evaluate whether patient-level and aneurysm-level clinical variables improved morphology-based prediction. The full early-fusion model jointly integrated all three information streams. Geometry and hemodynamic channels were processed through PointNeXt-based encoders, while clinical variables were processed through the clinical multilayer perceptron. The resulting geometry, flow, and clinical embeddings were concatenated and passed to a shared classification head. Early fusion allowed the classifier to learn cross-modal interactions during training, but it also required joint optimization of heterogeneous inputs with different scales, dimensions, and noise characteristics. All 735 known-status cases had a selected VTP geometry, generated PINN-derived hemodynamic descriptors, and sufficient clinical metadata after the stated missing-value handling.

### *2.4.3 Late-Fusion Rupture-Risk Scoring*

Late fusion was evaluated as a constrained probability-level ensemble. Rather than forcing all modalities into a single jointly optimized representation, late fusion combined the predictions of two independently trained models: Geometry + Clinical and Flow + Geometry. Using w for the Flow + Geometry contribution, the ensemble was defined as:

$$p_{late} = (1 - w)p_{geo+clin} + wp_{flow+geo}, \quad w = 0.30 \tag{7}$$

where $p_{geo+clin}$ denotes the case-level rupture-risk probability from the Geometry + Clinical model and $p_{flow+geo}$ denotes the probability from the Flow + Geometry model. The selected $w$ = 0.30 gives 70% weight to Geometry + Clinical and 30% to Flow + Geometry. This weighting preserved the stronger geometry-clinical prediction while incorporating complementary hemodynamic information as a probability-level correction. All reported rupture-risk comparisons used paired out-of-fold predictions from the same 735 labeled cases, so performance differences reflected input modality and fusion strategy rather than cohort composition.

To assess whether the fixed 70/30 weighting was sensitive to small changes in the mixing coefficient, the Flow + Geometry weight $w$ was swept from 0.00 to 1.00 in increments of 0.01, with Geometry + Clinical assigned the complementary weight 1 - w. Probability-space and logit-space blends were evaluated using the same 735 paired out-of-fold predictions. For each weight, pooled AUROC and AUPRC, mean held-out-fold AUROC, and worst-fold AUROC were computed. A stratified paired bootstrap with 5,000 resamples compared the selected 30% Flow + Geometry weight with Geometry + Clinical alone. The optimal weight was selected using the combined criteria of pooled discrimination, precision-recall performance, fold-level consistency, and model parsimony. Because the sweep used the same out-of-fold cohort, it was interpreted as a sensitivity analysis rather than independent hyperparameter validation.

## 3. Experimental Setting

### *3.1 Training Protocols*

All rupture-risk models used stratified five-fold cross-validation with a fixed random seed of 42. The same fold assignments were used across geometry-only, clinical-only, flow-augmented, and multimodal models to enable paired comparison of out-of-fold predictions. PointNet++ and the GNN used 4,096 sampled surface points, the selected PointNeXt geometry and multimodal models used 8,192 points, and the voxel CNN used a $24 \times 24 \times 24$ binary occupancy grid. Weighted sampling and label-smoothed losses addressed the 261:474 class imbalance. Model training was conducted on the NVIDIA A100 GPU cluster provided through Florida Institute of Technology's AI.Panther high-performance computing environment. Implementations used PyTorch, and metric calculation and feature-importance analysis used scikit-learn (Paszke et al., 2019; Pedregosa et al., 2011).

PointNet++, PointNeXt geometry-only, PointNeXt Flow + Geometry, Geometry + Clinical, and early-fusion models were trained for up to 220 epochs using AdamW with an initial learning rate of $3 \times 10^{-4}$, weight decay of $2 \times 10^{-4}$, batch size of 6, dropout probability of 0.30, and label-smoothed cross-entropy loss with smoothing factor 0.05. A one-cycle learning-rate schedule, mixed-precision training, gradient clipping at 1.0, and validation-AUROC early stopping with patience of 35 epochs were applied (Kingma & Ba, 2015; Loshchilov & Hutter, 2017, 2019; Smith, 2018; Szegedy et al., 2016). The voxel CNN used a $24 \times 24 \times 24$ occupancy grid, batch size of 4, initial learning rate of $2 \times 10^{-4}$, weight decay of $2 \times 10^{-4}$, and dropout probability of 0.25. The GNN used 4,096 sampled points, 24-nearest-neighbor graphs, three two-head graph-attention layers, a 128-dimensional hidden state, and a 256-dimensional case representation. GNN training used batch size 1 with four-step gradient accumulation, learning rate $3 \times 10^{-4}$, weight decay $3 \times 10^{-4}$, dropout probability 0.25, focal-loss parameter $\gamma = 1.5$, label smoothing 0.03, gradient clipping at 1.0, and validation-AUROC early stopping with patience of 45 epochs over at most 250 epochs. Gradient checkpointing and chunked nearest-neighbor construction were used to control memory usage.

For aneurysm detection, the PointNeXt detector described in Section 2.2 was trained on the VMR cohort using stratified five-fold cross-validation. Each geometry was represented by 2,048 sampled surface points. Random-start farthest-point sampling was used during training to increase sampling variability, whereas deterministic sampling was used during evaluation to ensure reproducible predictions. Class imbalance between aneurysm-positive and non-aneurysm geometries was addressed using focal loss and fold-specific class weights (Lin et al., 2017).

For each aneurysm case, the PINN was trained independently using the formulation described in Section 2.3. The implementation used a residual multilayer perceptron with five 128-unit SiLU blocks, skip connections, and 16 Fourier-feature modes, as illustrated in **Figure S2**. In the reported implementation, each optimization run retained 1,024 interior points, 512 wall points, 22 inlet points, and 32 outlet points, with space-time collocation points resampled during optimization. The network was trained for 1,000 epochs using Adam with an initial learning rate of 0.001, cosine warm restarts at epochs 400 and 800, and gradient clipping at 1.0. The modeled cardiac-cycle duration was 1.0 s, blood density was 1,060 kg/m³, dynamic viscosity was 0.0035 Pa·s, and the prescribed mean inlet flow rate was 0.2 mL/s.

AneuX coordinates were stored in millimeters. Spatial coordinates and time were nondimensionalized using case-specific length and convective time scales before network evaluation. Exported velocity was restored using the case-specific inlet velocity scale and reported in m/s. Relative pressure was scaled by $\rho u^2$ and reported in Pa when the physical scale was retained; the representative pressure field in **Figure 4** is shown as normalized relative pressure because the saved field was gauge-normalized. Velocity gradients were converted from inverse millimeters to inverse meters before multiplication by dynamic viscosity, yielding instantaneous WSS and TAWSS in Pa. OSI is dimensionless, and RRT is reported in $Pa^{-1}$.

### *3.2 Evaluation Metrics and Statistical Testing*

Threshold-independent and threshold-dependent metrics were computed from pooled out-of-fold predictions. AUROC and AUPRC were the primary comparison metrics. Threshold-dependent metrics included accuracy, precision, sensitivity, specificity, balanced accuracy, F1 score, and confusion matrices. Model-specific thresholds were selected to maximize pooled out-of-fold F1 score. For each selected threshold, accuracy, precision, sensitivity, specificity, and F1 were also recomputed separately in each held-out fold; their sample standard deviations quantify fold-to-fold operating-point variability (Davis & Goadrich, 2006; Fawcett, 2006; Saito & Rehmsmeier, 2015).

Rupture-risk metrics were computed using the 735 AneuX cases with documented rupture status. For aneurysm detection, the positive class was defined as aneurysm-positive geometry. For rupture-risk prediction, the positive class was defined as ruptured aneurysm. Paired AUROC differences were evaluated using two-sided DeLong tests because every rupture-risk model produced an out-of-fold probability for the same 735 labeled cases. Holm correction was used to control the familywise error rate within the geometry-only and multimodal comparison families. Adjusted $p < 0.05$ was considered statistically significant (DeLong et al., 1988). Statistical comparisons were interpreted together with pooled AUROC, pooled AUPRC, and fold-level performance trends.

### *3.3 Explainability and Feature-Importance Analysis*

A separate feature-importance analysis was performed to assess rupture-risk-associated geometric, hemodynamic, and clinical descriptors at the tabular feature level. This analysis was complementary to the point-cloud models. It identified descriptors associated with the observed labels, but was not used to infer or explain the internal decision process of the PointNeXt encoders. Each labeled aneurysm was converted into one tabular feature row. Geometry descriptors included bounding-box, radial, principal-component, and convex-hull measures. Hemodynamic descriptors summarized PINN-derived TAWSS, OSI, RRT, threshold fractions, interaction terms, and field correlations. Clinical descriptors included age, biological sex, aneurysm location, and side. Age and biological sex were encoded numerically, while aneurysm location and side were one-hot encoded. Feature importance was evaluated using five patient-aware folds. Within each training fold, missing-value imputation, feature standardization, and categorical encoding were fitted using training-fold data only and then applied to the held-out fold. A class-balanced regularized logistic-regression model was trained and evaluated using the same five-fold structure.

Feature importance was quantified using a composite score that combined two measures: the absolute standardized logistic-regression coefficient and the decrease in held-out AUROC after single-feature permutation. Both measures were converted to percentile ranks and averaged for each descriptor. Modality-level importance was computed as the mean composite score across retained descriptors within each modality, reducing bias from unequal feature counts. Redundant shear-derived variables and raw RRT summaries were excluded from this analysis.

## 4. Results

The results are organized according to the major components of the proposed framework. Section 4.1 reports aneurysm detection on the VMR cohort. Section 4.2 summarizes PINN optimization behavior and representative PINN-derived hemodynamic descriptors. Sections 4.3 and 4.4 present geometry-only and multimodal rupture-risk prediction performance. Section 4.5 reports descriptor-level feature-importance analysis.

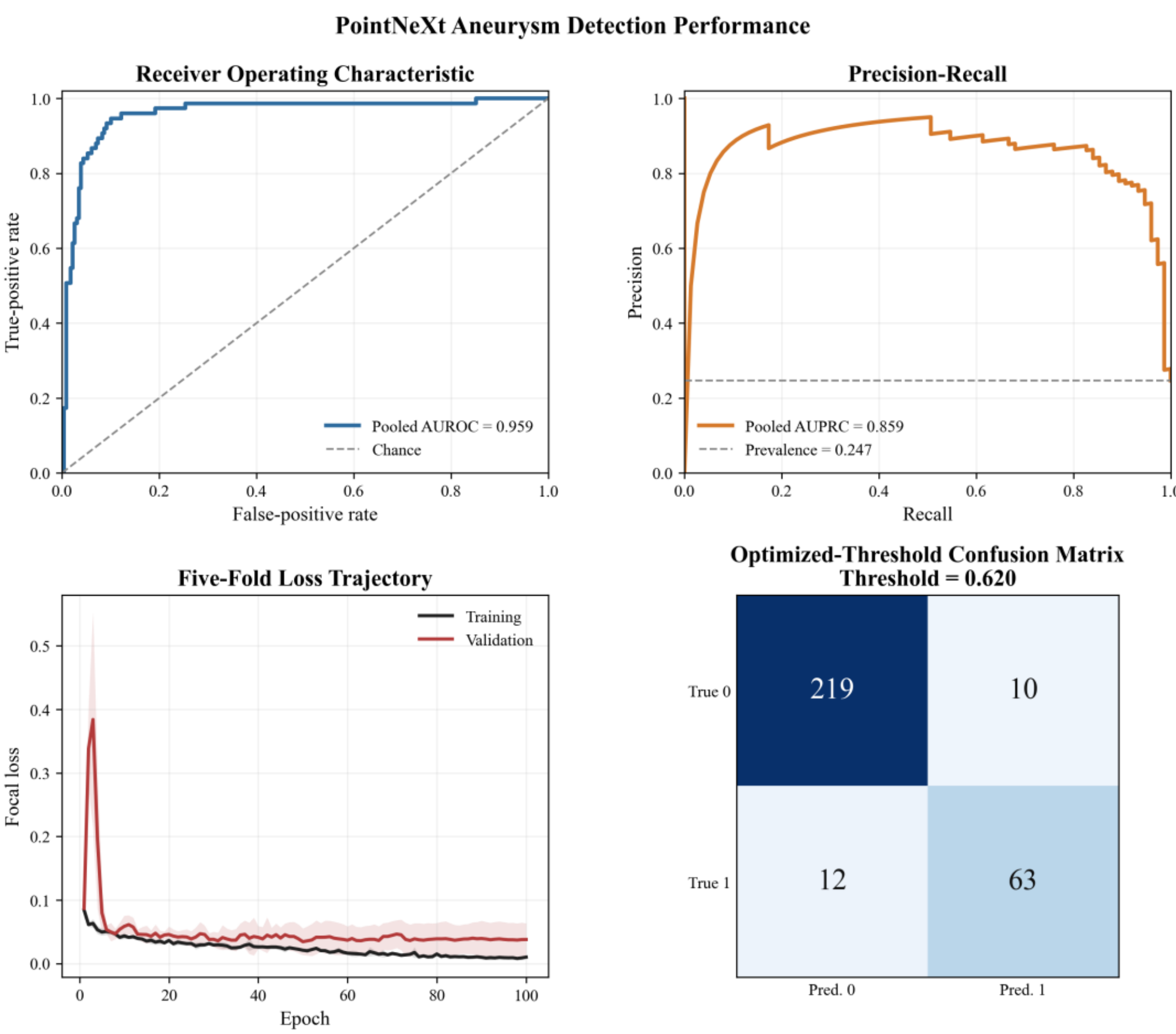

**Figure 2. PointNeXt aneurysm-detection performance on the VMR cohort**. Pooled out-of-fold ROC and precision-recall curves are shown across 304 vascular geometries, with AUROC 0.959 and AUPRC 0.859. The dashed line in the precision-recall panel denotes aneurysm prevalence. The loss panel shows mean training and validation focal loss across five folds, with shaded bands indicating one standard deviation. The confusion matrix is reported at the F1-optimized threshold of 0.620, yielding 219 true negatives, 10 false positives, 12 false negatives, and 63 true positives.

### *4.1 Aneurysm Detection Performance*

The PointNeXt aneurysm detector achieved strong internal performance on the 304-case VMR cohort. Using pooled out-of-fold predictions from stratified five-fold cross-validation, the detector achieved AUROC 0.959 and AUPRC 0.859 for aneurysm-positive versus non-aneurysm classification. Across the five held-out folds, the mean best-checkpoint AUROC was 0.963 +/- 0.028, and the mean AUPRC was 0.888 +/- 0.087. At the F1-optimized threshold of 0.620, the detector achieved accuracy 0.928, precision 0.863, recall 0.840, and F1 score 0.851. The pooled confusion matrix included 219 true negatives, 10 false positives, 12 false negatives, and 63 true positives. These results show that the PointNeXt detector learned vascular shape patterns associated with aneurysm presence in the internal cross-validation setting. The pooled ROC curve, precision-recall curve, training and validation loss trajectories, and confusion matrix are shown in **Figure 2**.

### *4.2 PINN Training Behavior and Hemodynamic Descriptor Generation*

The unsteady PINN was trained independently for each aneurysm geometry to generate geometry-conditioned, physics-constrained hemodynamic descriptors. In the representative case, the PINN completed 1,000 recorded epochs. The weighted optimization loss decreased from 192.464 at epoch 1 to 0.002800 at epoch 1,000 and reached its minimum value of 0.002678 at epoch 988. Scheduler restarts produced short transient increases, but the overall loss envelope continued to decline.

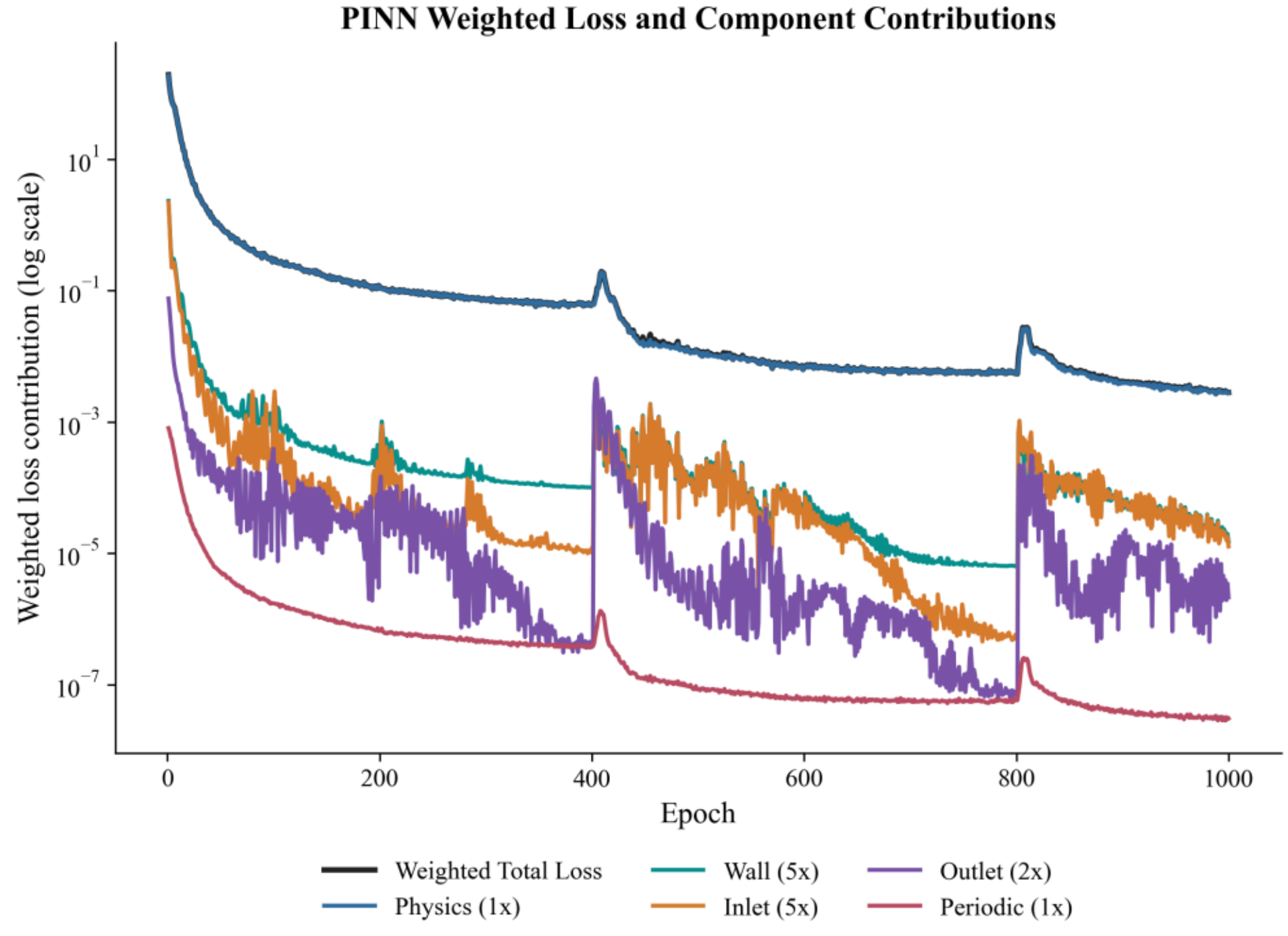

**Figure 3. Weighted PINN optimization loss for a representative aneurysm case over 1,000 epochs**. The weighted total loss combines the Navier–Stokes physics residual, wall no-slip loss, inlet velocity loss, outlet pressure-gauge loss, and periodicity loss. The logarithmic scale shows loss components spanning several orders of magnitude.

The loss trajectories are shown in **Figure 3**. The weighted optimization loss combined the Navier-Stokes physics residual, wall no-slip loss, inlet velocity loss, outlet pressure-gauge loss, and periodicity loss. The decreasing loss indicates improved satisfaction of the prescribed residual and boundary-condition constraints during optimization. Because no paired CFD or in vivo flow measurements were used as supervised targets, the loss trajectories are interpreted as evidence of optimization convergence rather than independent validation of physiological flow accuracy. Representative PINN-derived fields are shown in **Figure 4** for ANSYS_UNIGE_09_cut1 at t = 0.22 s. The top row shows transient relative pressure, velocity magnitude, and instantaneous WSS at that time point. The bottom row shows time-aggregated TAWSS, OSI, and RRT over the modeled cardiac cycle. Instantaneous WSS and TAWSS use the same color scale. Their spatial patterns are very similar in this representative case (pointwise Pearson correlation 0.998) because the modeled WSS distribution changed little over the cardiac cycle; TAWSS is the cycle average. These outputs are interpreted as model-derived, physics-constrained hemodynamic descriptors under the stated assumptions.

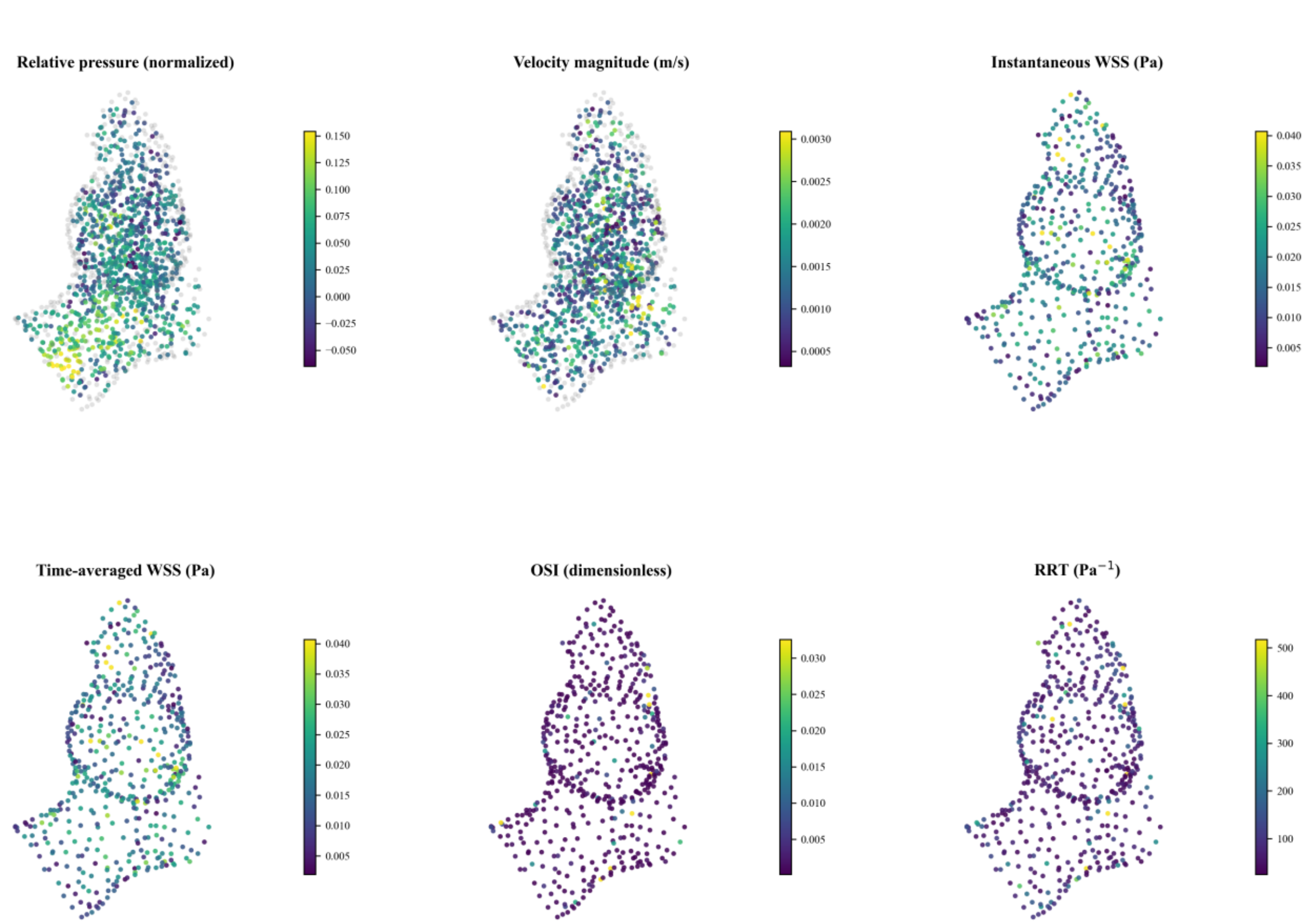

**Figure 4. Representative PINN-derived fields for ANSYS_UNIGE_09_cut1**. The top row shows normalized transient relative pressure, velocity magnitude in m/s, and instantaneous WSS in Pa at t = 0.22 s. The bottom row shows time-aggregated TAWSS in Pa, dimensionless OSI, and RRT in $Pa^{-1}$ over the modeled cardiac cycle. Instantaneous WSS and TAWSS share one color scale. Their similar maps reflect a pointwise correlation of 0.998 in this case, rather than separate panel normalization.

### *4.3 Geometry-Only Rupture-Risk Prediction*

Geometry-only models were evaluated to determine the extent to which vascular morphology alone discriminated ruptured from unruptured aneurysms. Four geometry representations were compared on the same 735 AneuX cases with documented rupture status: PointNet++, PointNeXt, voxel CNN, and GNN. Overall, geometry-only models showed measurable but limited discrimination. PointNeXt achieved the highest pooled AUROC among geometry-only models at 0.611, followed by PointNet++ at 0.590, voxel CNN at 0.564, and GNN at 0.517. In contrast, PointNet++ achieved the highest pooled AUPRC at 0.459, followed by PointNeXt at 0.448, voxel CNN at 0.420, and GNN at 0.372. These results indicate that point-cloud representations captured more useful geometric information than voxel- or graph-based baselines, although the overall performance remained moderate and insufficient for strong rupture-risk prediction using morphology alone.

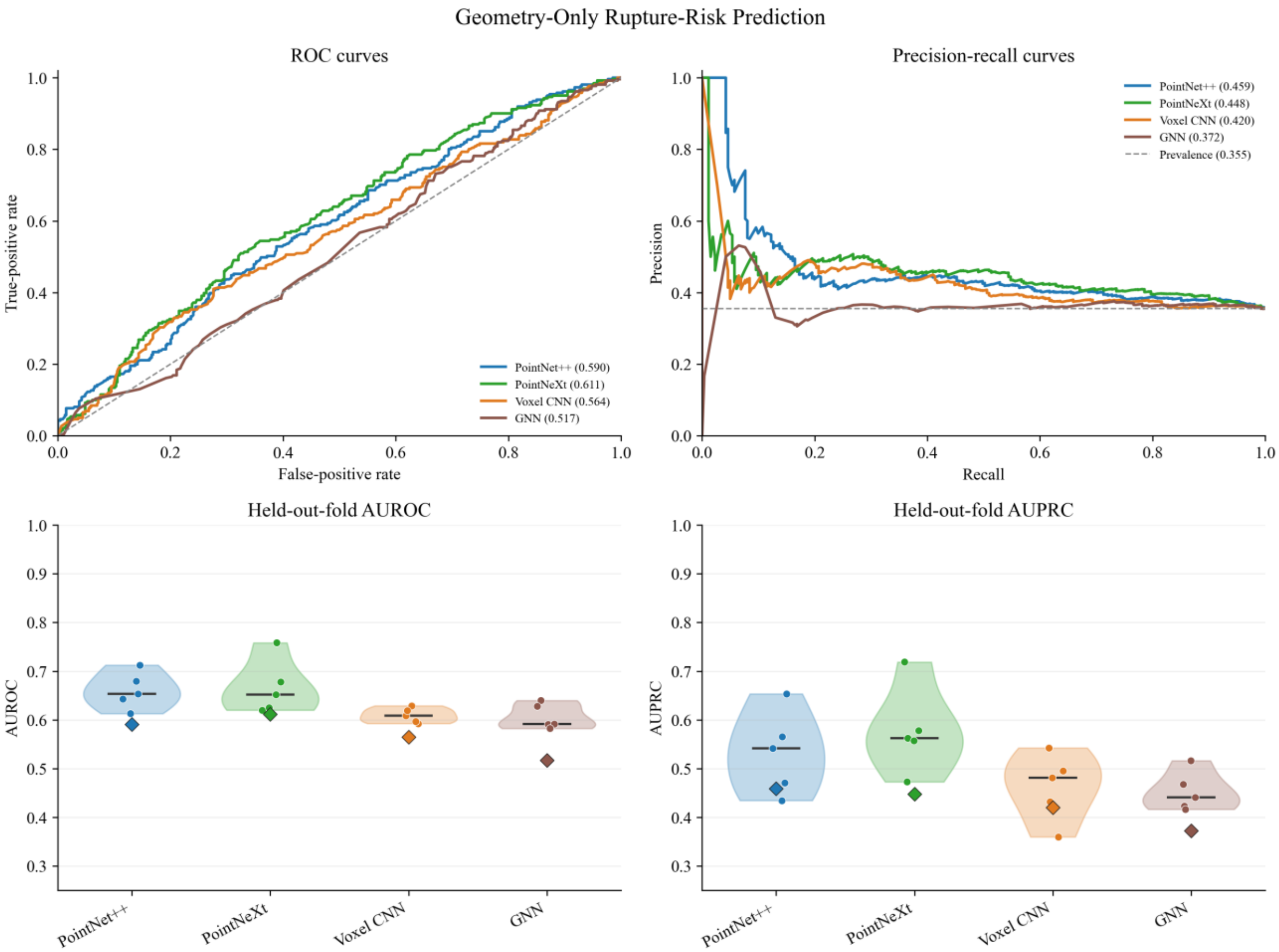

**Figure 5. Geometry-only rupture-risk prediction performance**. Pooled out-of-fold ROC and precision-recall curves are shown for PointNet++, PointNeXt, voxel CNN, and GNN on the 735-case AneuX rupture-risk cohort. The dashed precision-recall baseline denotes rupture prevalence. Bottom panels show held-out-fold AUROC and AUPRC distributions; circles denote individual folds, and diamonds denote pooled out-of-fold estimates.

PointNeXt also achieved the highest mean held-out fold AUROC/AUPRC, 0.667/0.578. However, the pooled AUROC difference between PointNeXt and PointNet++ was not significant after Holm correction. PointNet++ also did not differ significantly from the voxel CNN after correction, whereas the PointNet++ versus GNN comparison reached significance in the reported paired analysis. These results indicate that point-cloud models captured useful vascular morphology, but geometry alone was insufficient for strong rupture-risk prediction. Geometry-only ROC curves and fold-level AUROC and AUPRC distributions are combined in **Figure 5**, and pooled confusion matrices are shown in **Figure S4**. Full threshold-independent and optimized-threshold metrics are summarized in **Table 1**.

| Model | AUROC | AUPRC | Thr. | Acc. | Prec. | Recall | Spec. | F1 |
|---|---|---|---|---|---|---|---|---|
| PointNet++ | 0.590 (0.038) | 0.459 (0.086) | 0.413 | 0.445 (0.123) | 0.383 (0.076) | 0.920 (0.082) | 0.184 (0.203) | 0.541 (0.063) |
| **PointNeXt** | **0.611 (0.056)** | **0.448 (0.089)** | **0.446** | **0.468 (0.093)** | **0.392 (0.048)** | **0.900 (0.083)** | **0.230 (0.166)** | **0.546 (0.050)** |
| Voxel CNN | 0.564 (0.015) | 0.420 (0.070) | 0.025 | 0.384 (0.082) | 0.363 (0.049) | 0.977 (0.051) | 0.057 (0.134) | 0.530 (0.046) |
| GNN | 0.517 (0.026) | 0.372 (0.041) | 0.484 | 0.393 (0.071) | 0.365 (0.033) | 0.962 (0.086) | 0.080 (0.159) | 0.530 (0.037) |

**Table 1. Geometry-only rupture-risk prediction performance among 735 labeled AneuX cases.** AUROC and AUPRC are pooled out-of-fold estimates, with sample standard deviations across five held-out folds shown in parentheses. Threshold-dependent metrics were computed at the model-specific F1-optimized threshold shown in the Thr. column.

### *4.4 Multimodal Rupture-Risk Prediction*

Multimodal models were evaluated to determine whether clinical variables and PINN-derived hemodynamic descriptors improved rupture-risk prediction beyond geometry-only learning. All models used the same 735 labeled cases and identical fold assignments to enable paired comparison of out-of-fold predictions.

The clinical-only model achieved pooled AUROC/AUPRC of 0.784/0.632, substantially exceeding the PointNeXt geometry-only model, which achieved 0.611/0.448. Adding PINN-derived hemodynamic descriptors to geometry improved performance to 0.738/0.582 for the Flow + Geometry model. Combining geometry with clinical variables further improved performance to 0.809/0.701. The early-fusion model, which jointly integrated geometry, hemodynamic descriptors, and clinical variables in a single representation, achieved 0.803/0.676.

The fixed 70/30 late-fusion model achieved the highest overall performance, with pooled AUROC 0.827 and AUPRC 0.732. In Holm-corrected paired two-sided DeLong tests, late fusion exceeded the clinical-only model (adjusted $p = 0.0116$), PointNeXt geometry (adjusted $p < 0.001$), Flow + Geometry (adjusted $p < 0.001$), Geometry + Clinical (adjusted $p = 0.00590$), and early fusion (adjusted $p = 0.0259$). Overall, the largest performance gain came from adding clinical variables to geometry, while PINN-derived hemodynamic descriptors provided additional complementary improvement through Flow + Geometry and late fusion.

Late-fusion weight sensitivity identified the probability-space 70/30 blend as the selected setting, with pooled AUROC 0.8272, AUPRC 0.7322, mean held-out-fold AUROC 0.8454, and worst-fold AUROC 0.7987. The selected 30% Flow + Geometry weight also lay within a stable 25%-35% performance plateau, supporting the rounded coefficient as a parsimonious choice rather than a uniquely optimal constant. Pooled ROC and precision-recall curves with fold-level AUROC and AUPRC distributions are combined in **Figure 6**. Confusion matrices, late-fusion threshold sensitivity, and late-fusion weight sensitivity are shown in **Figures S5**, **7**, and **8**, respectively. Full multimodal performance metrics are summarized in **Table 2**.

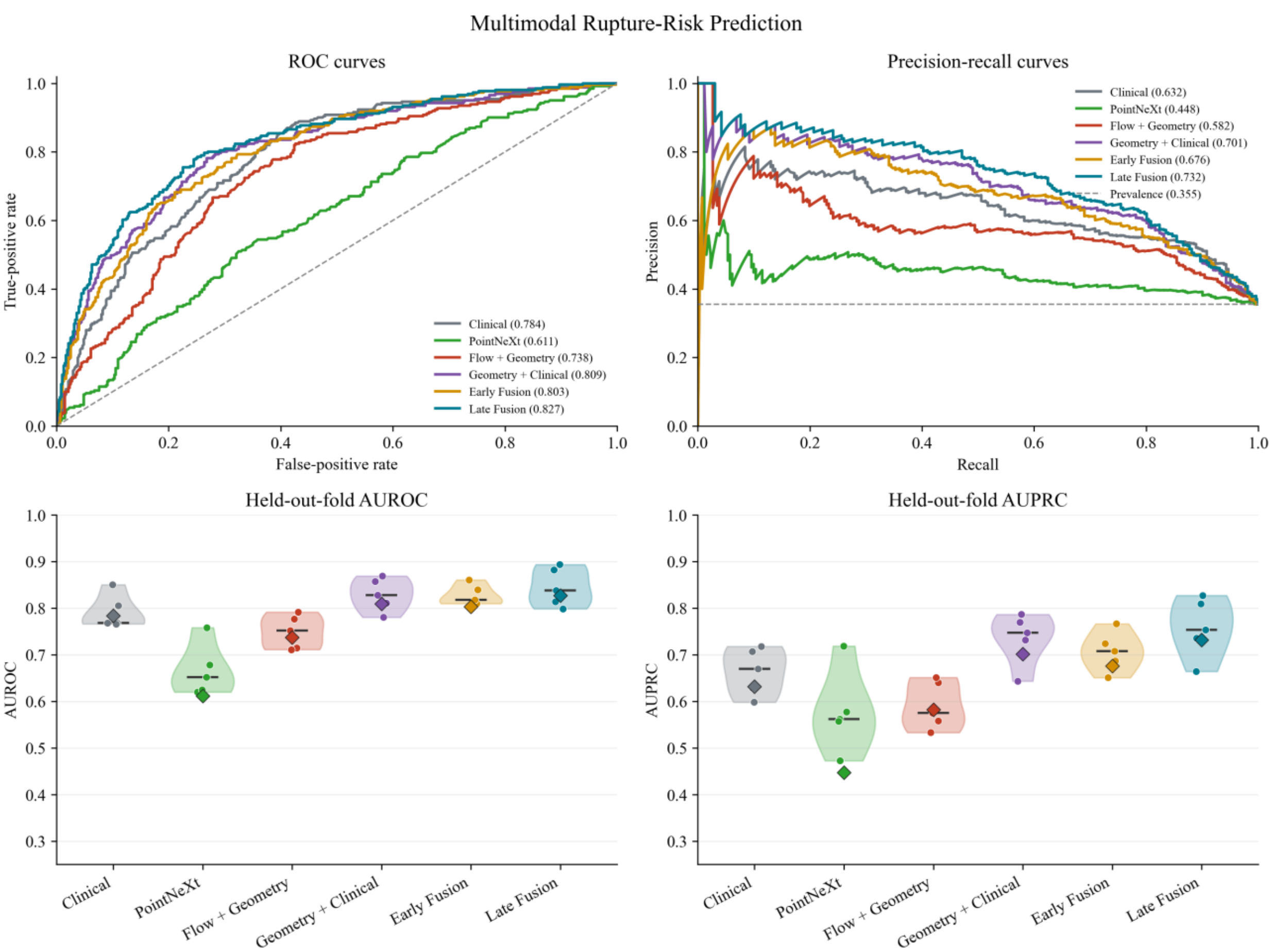

**Figure 6. Multimodal rupture-risk prediction performance**. Top panels show pooled out-of-fold ROC and precision-recall curves for clinical-only, PointNeXt geometry, Flow + Geometry, Geometry + Clinical, early fusion, and fixed 70/30 late fusion. Bottom panels show held-out-fold AUROC and AUPRC distributions; circles denote folds and diamonds denote pooled estimates.

| Model | AUROC | AUPRC | Thr. | Acc. | Prec. | Recall | Spec. | F1 |
|---|---|---|---|---|---|---|---|---|
| Clinical | 0.784 (0.037) | 0.632 (0.051) | 0.480 | 0.684 (0.038) | 0.533 (0.036) | 0.885 (0.061) | 0.574 (0.075) | 0.666 (0.025) |
| PointNeXt Geometry | 0.611 (0.056) | 0.448 (0.089) | 0.446 | 0.468 (0.093) | 0.392 (0.048) | 0.900 (0.083) | 0.230 (0.166) | 0.546 (0.050) |
| Flow + Geometry | 0.738 (0.036) | 0.582 (0.052) | 0.483 | 0.663 (0.054) | 0.516 (0.046) | 0.824 (0.125) | 0.574 (0.135) | 0.634 (0.033) |
| Geometry + Clinical | 0.809 (0.036) | 0.701 (0.056) | 0.504 | 0.750 (0.021) | 0.616 (0.080) | 0.782 (0.185) | 0.732 (0.110) | 0.689 (0.063) |
| Early Fusion | 0.803 (0.022) | 0.676 (0.043) | 0.472 | 0.718 (0.042) | 0.575 (0.081) | 0.793 (0.158) | 0.677 (0.136) | 0.667 (0.043) |
| Late Fusion | **0.827 (0.042)** | **0.732 (0.065)** | **0.510** | **0.762 (0.025)** | **0.633 (0.051)** | **0.785 (0.170)** | **0.749 (0.085)** | **0.701 (0.066)** |

**Table 2. Multimodal rupture-risk prediction performance.** AUROC and AUPRC are pooled out-of-fold estimates, with sample standard deviations across five held-out folds shown in parentheses. Threshold-dependent metrics were computed at the model-specific F1-optimized threshold shown in the Thr. column. Late Fusion denotes the fixed 70/30 probability-level ensemble of Geometry + Clinical and Flow + Geometry models.

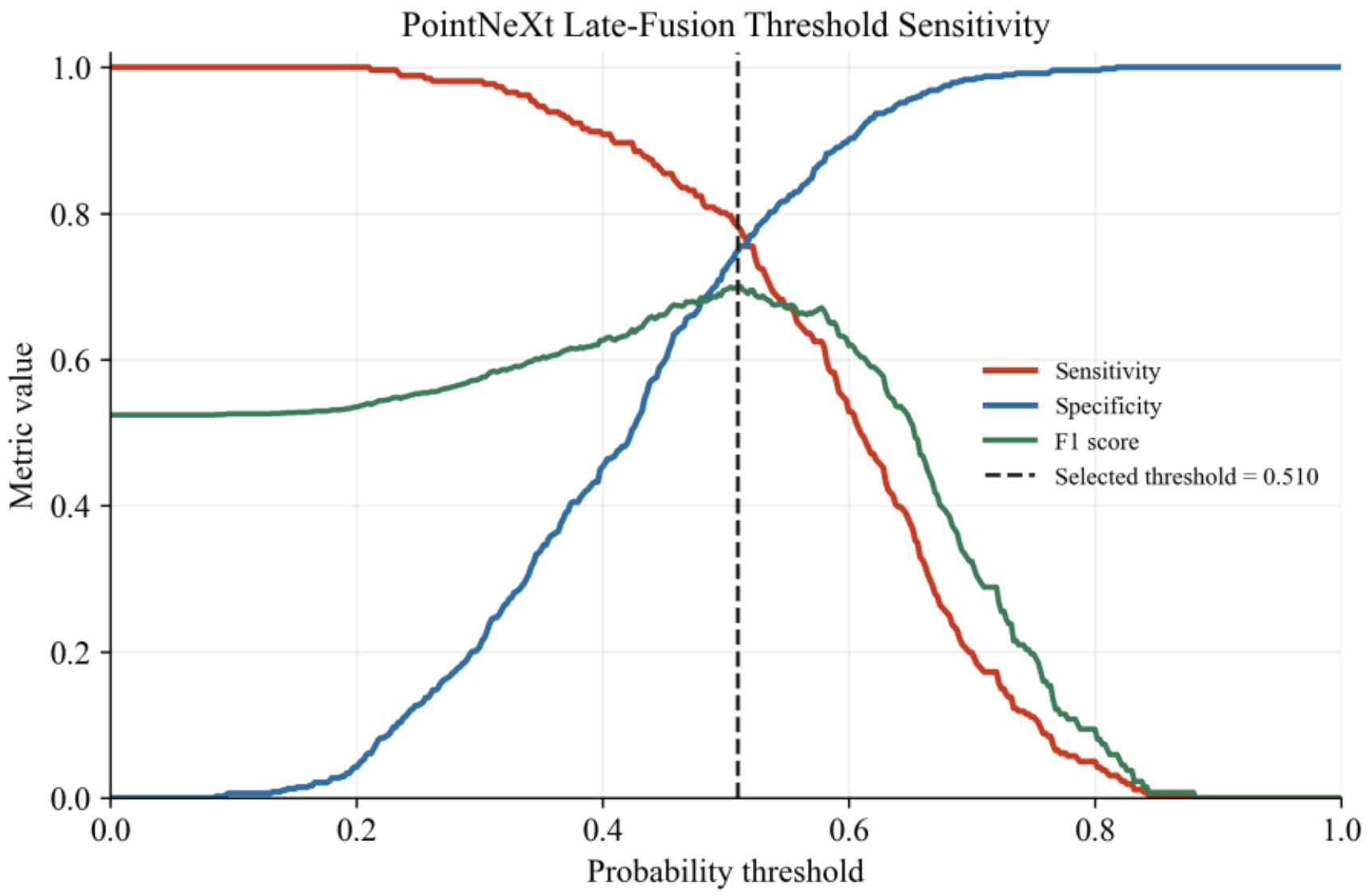

**Figure 7. Late-fusion threshold sensitivity**. Sensitivity, specificity, and F1 score are shown across probability thresholds for the fixed 70/30 late-fusion rupture-risk model. The dashed vertical line denotes the F1-optimized threshold of 0.510, where the model balances sensitivity and specificity while maximizing F1 score.

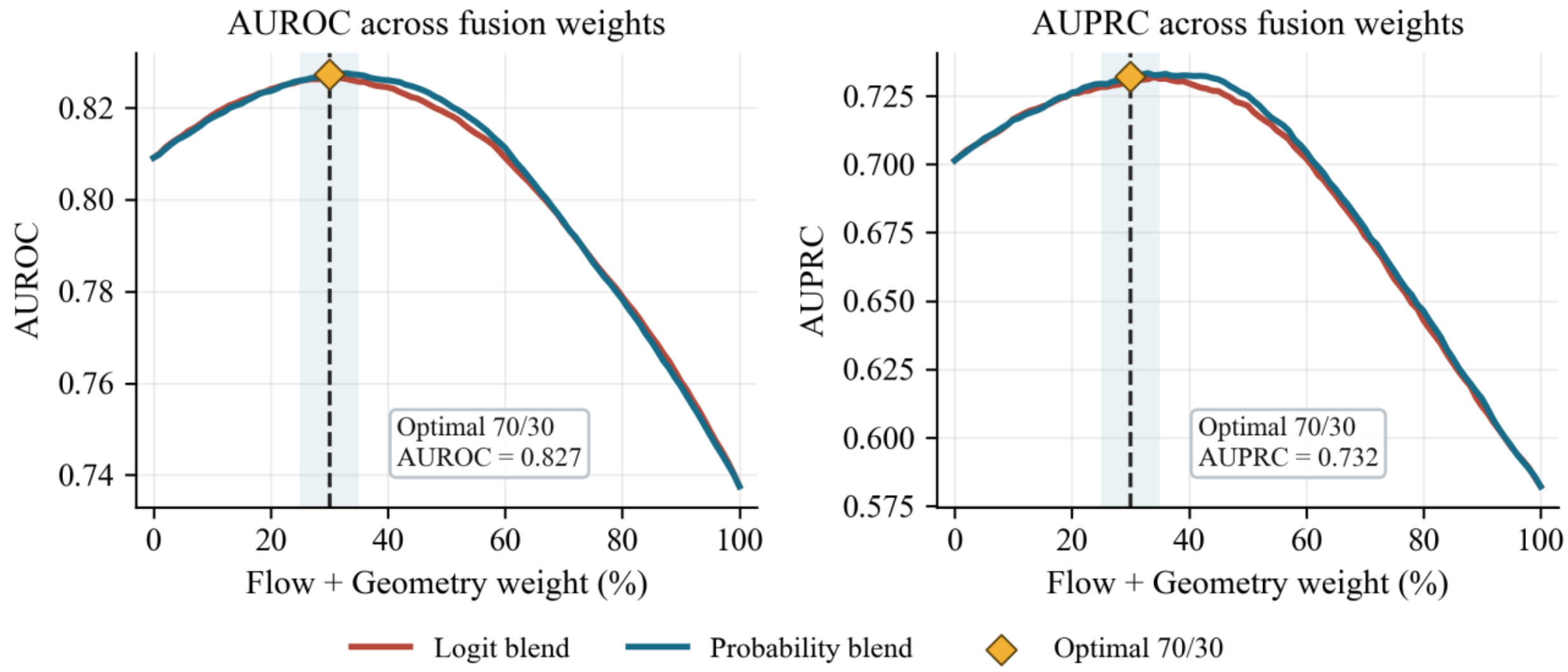

**Figure 8. Late-fusion weight sensitivity**. Pooled out-of-fold AUROC (left) and AUPRC (right) are shown across Flow + Geometry weights from 0% to 100%, with Geometry + Clinical assigned the complementary weight. Curves compare probability- and logit-space blending on the same 735 labeled cases. The dashed line marks the selected 30% Flow + Geometry weight, and the shaded region marks the stable 25%-35% performance plateau.

### *4.5 Rupture-Risk Feature Importance*

Feature-importance analysis was recomputed on all 735 aneurysms with documented rupture status using the restored composite-ranking method and the retained interpretable descriptors. The regularized tabular model achieved mean AUROC 0.774 +/- 0.026 and mean AUPRC 0.617 +/- 0.047 across five held-out folds. Mean composite importance per retained feature was 0.600 for geometry, 0.518 for hemodynamics, and 0.441 for clinical descriptors. The three modalities therefore showed more comparable contributions, with geometry ranking first and hemodynamics close behind after adjustment for the unequal numbers of retained descriptors.

At the descriptor level, the 75th percentile of OSI had the highest composite score (0.985), followed by cavernous internal carotid artery location (0.977), TAWSS standard deviation (0.954), anterior communicating artery location (0.954), radial-distance standard deviation (0.946), posterior communicating internal carotid artery location (0.938), and radial-distance mean (0.915). These rankings combine coefficient magnitude with held-out permutation behavior and should be interpreted as relative predictive importance rather than causal effects. The revised feature- and modality-level results are shown in **Figure 9.**

## 5. Discussion

### *5.1 Principal Findings and Interpretation*

This study developed a staged framework for cerebral aneurysm detection followed by rupture-risk prediction. The key contribution is a change in supervision and information flow relative to recent literature. Current CTA, radiomics, point-cloud, and morphology-based systems generally

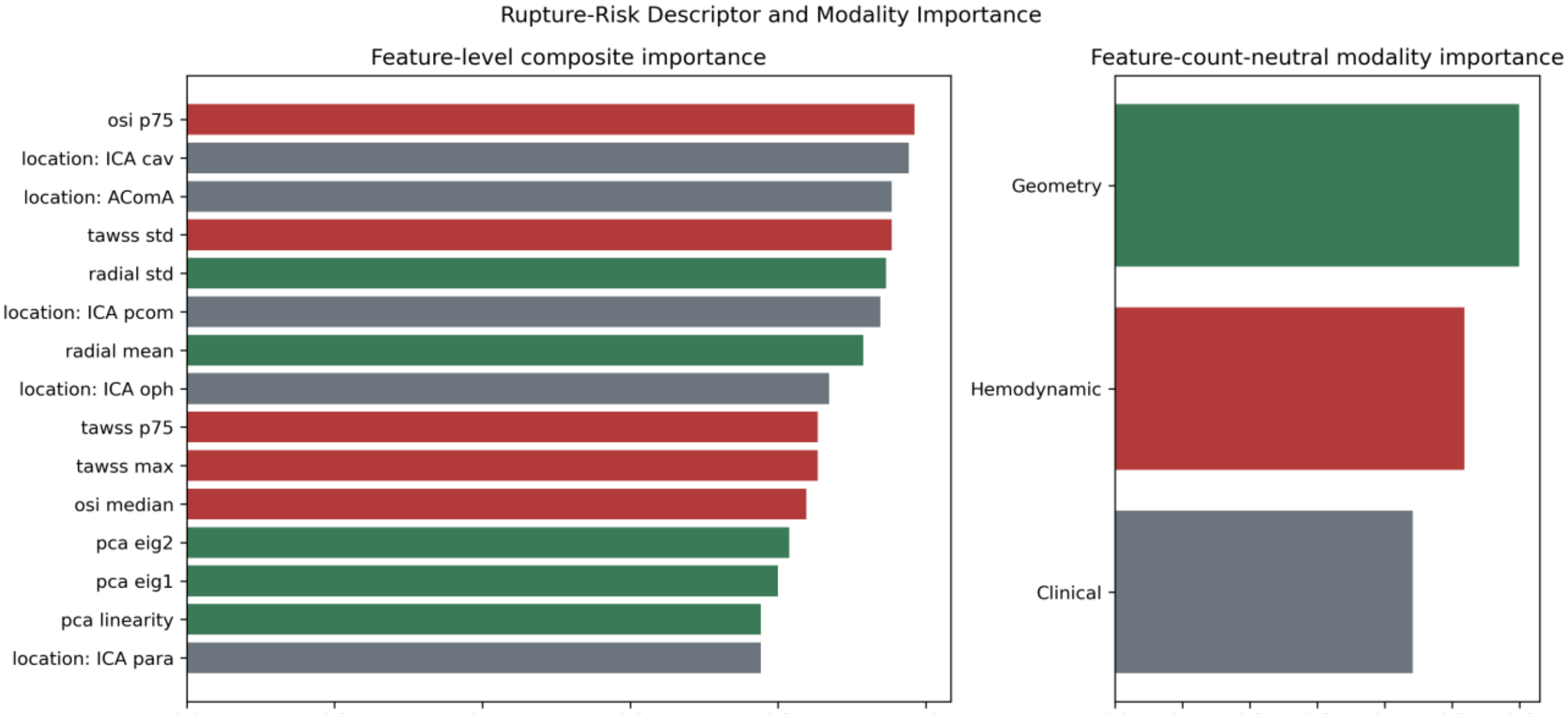

**Figure 9. Composite descriptor and modality importance**. Left: the 15 highest feature-level scores obtained by averaging the percentile ranks of absolute standardized coefficients and held-out single-feature permutation importance. Right: mean composite importance per retained descriptor for geometry, hemodynamic, and clinical modalities. Averaging within each modality adjusts for unequal feature counts.

learn rupture-associated appearance or shape directly (Cao et al., 2024; Johnson et al., 2025; Kofman et al., 2026; Li et al., 2023; Luo et al., 2023; Shi et al., 2026; Zeng et al., 2025), while recent hemodynamic field-prediction models learn CFD-generated fields from paired simulations (Rajhi et al., 2025; Sheng et al., 2026). Here, PointNeXt first performs aneurysm detection; for aneurysm-positive cases, a PINN generates physics-constrained pressure, velocity, instantaneous WSS, average WSS (TAWSS), OSI, and RRT descriptors without paired CFD field labels; and separate geometry, flow, and clinical branches are compared through early and probability-level late fusion on identical out-of-fold cases. The fixed 70/30 late-fusion model achieved the strongest rupture-risk discrimination among evaluated models. The results show that clinical variables provided the strongest individual signal, while PINN-derived hemodynamic descriptors added complementary information that improved performance when combined with geometry and clinical context

The PINN module should be interpreted as a physics-constrained descriptor generator. Decreasing physics and boundary-condition losses indicate improved satisfaction of the imposed equations but do not establish physiological accuracy. Feature-count-neutral composite importance placed geometry first, followed closely by hemodynamics, while clinical descriptors retained a comparable but lower mean score. Leading individual signals included the OSI distribution, aneurysm location, radial geometry, and TAWSS magnitude. These tabular rankings do not directly explain the internal PointNeXt representations and should not be interpreted causally.

### *5.2 Limitations*

This study has several limitations. First, the rupture-risk models were trained using cross-sectional ruptured-versus-unruptured labels. They produce case-level risk scores but do not estimate the timing or probability of a future rupture for an individual unruptured aneurysm; longitudinal cohorts with serial imaging and prospective follow-up are needed for that purpose. Second, all prediction results were obtained using internal cross-validation on open datasets, and external validation is needed to assess generalization across institutions, imaging protocols, segmentation pipelines, and patient populations. Third, the PINN-derived hemodynamic descriptors rely on simplified assumptions, including rigid walls, incompressible Newtonian flow, a prescribed inlet waveform, and outlet pressure-gauge specification. Patient-specific inlet and outlet boundary conditions were not available, and the generated flow fields were not validated against conventional CFD or in vivo flow measurements.

### *5.3 Future Work*

Future work will be focusing on external validation, hemodynamic verification, and longitudinal modeling. Independent multicenter cohorts are needed to evaluate the robustness of aneurysm detection and rupture-risk prediction across clinical imaging and segmentation workflows. PINN-derived fields should be compared with patient-specific CFD simulations and, when available, in vivo flow measurements to quantify agreement in pressure, velocity, WSS, OSI, and RRT. Incorporating patient-specific inflow and outflow conditions may further improve the physiological relevance of these descriptors. Learning or adaptively balancing the PINN loss weights will also be investigated as an alternative to fixed empirically tuned values. Longitudinal follow-up is essential because prospective rupture-risk modeling requires serial imaging and future outcome data. This vascular modeling framework could also be extended to evaluate aneurysm growth, post-treatment hemodynamic changes, and response after coils, flow diverters, or other endovascular interventions.

## 6. Conclusion

This study presented a modular framework for cerebral aneurysm detection and rupture-risk prediction by integrating three-dimensional vascular geometry learning, physics-informed hemodynamic descriptors, and clinical variables. The PointNeXt-based detector achieved pooled out-of-fold AUROC of 0.959 and AUPRC of 0.859. For rupture-risk prediction, geometry-only models showed limited discrimination, whereas multimodal integration improved performance. The fixed 70/30 late-fusion model achieved the highest overall performance, with pooled AUROC of 0.827 and AUPRC of 0.732. Feature-count-neutral composite importance showed comparable geometry, hemodynamic, and clinical contributions, with OSI distribution, aneurysm location, radial geometry, and TAWSS among the leading descriptors to cross-sectional rupture-risk discrimination. Overall, the proposed framework provides a scalable and interpretable basis for integrating vascular morphology, physics-constrained flow descriptors, and clinical information in aneurysm assessment. Future validation on independent multicenter and longitudinal cohorts will be important for establishing its utility in prospective rupture-risk modeling and treatment-response analysis.

**Acknowledgements**

The authors acknowledge the Regeneron International Science and Engineering Fair (ISEF) for providing a forum for the original presentation of this research where it received 1st Place in Computational Biology & Bioinformatics. X. Li's work was partially funded by the Division of Mathematical Sciences, National Science Foundation with the award numbers 2410678.

**Declaration of generative AI and AI-assisted technologies in the manuscript preparation process**

During the preparation of this work, the authors used ChatGPT to improve language and readability and to assist with organizing flowchart text and layout. The authors reviewed, edited, and verified the resulting manuscript text and manually finalized the submitted figures in conventional presentation and plotting software. The authors take full responsibility for the content of the published article.

## Supplementary Information

### PointNeXt Aneurysm Detection Architecture

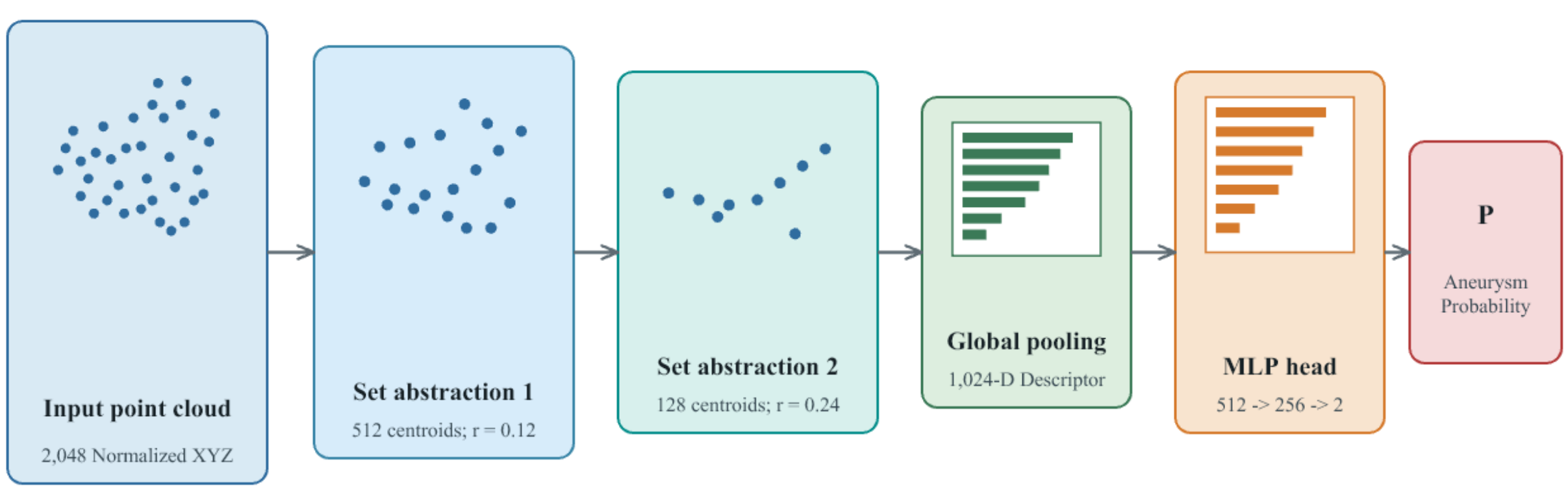

**Supplementary Figure S1. PointNeXt architecture used for aneurysm detection**. A normalized 2,048-point vascular surface cloud is processed by two hierarchical set-abstraction stages containing 512 and 128 centroids, respectively. Global pooling produces a 1,024-dimensional case representation, and a multilayer perceptron maps this representation to the aneurysm-presence probability.

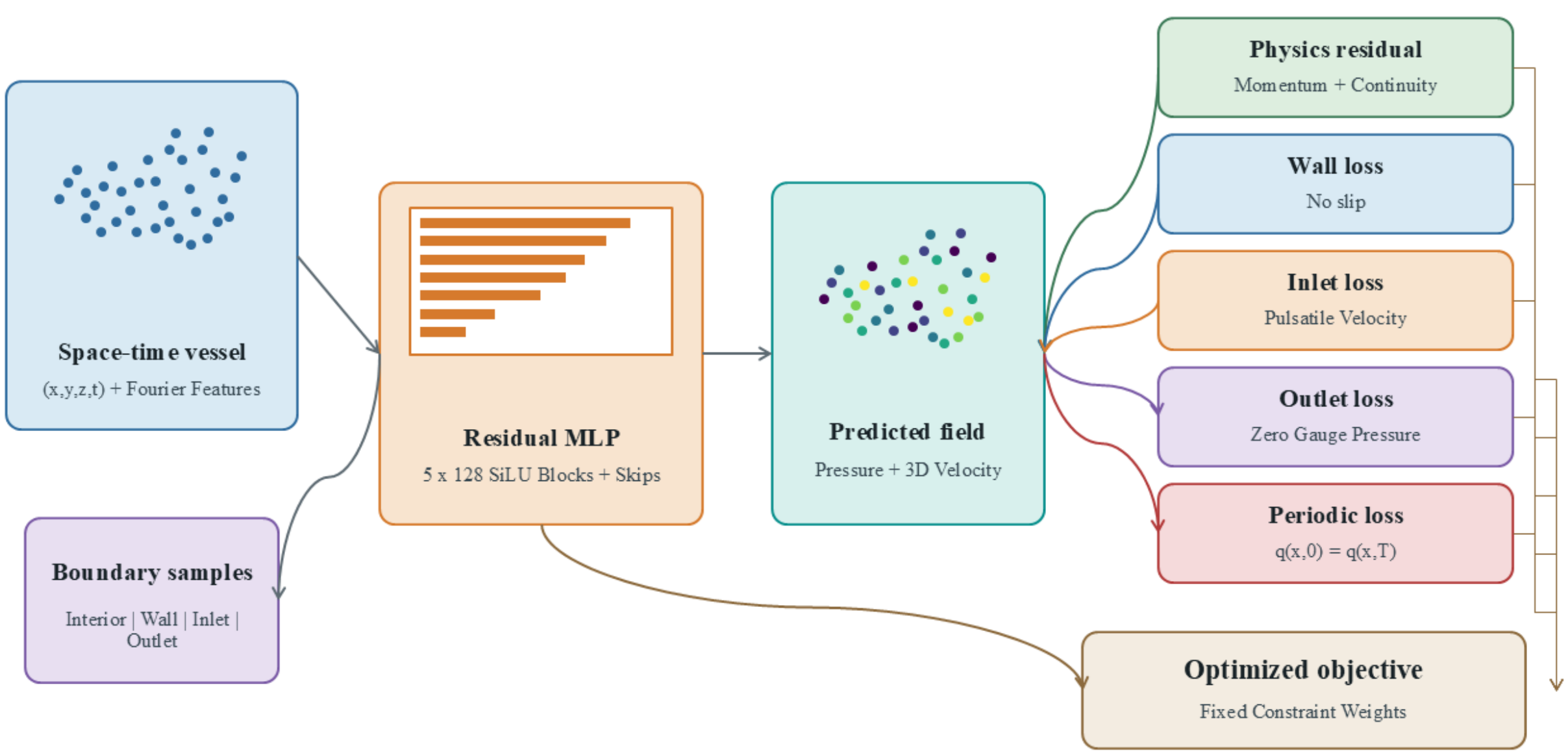

**Supplementary Figure S2. Unsteady PINN framework for hemodynamic field estimation**. Sampled vascular geometry and boundary points are combined with Fourier-encoded space-time coordinates and passed through a residual multilayer perceptron. The network predicts relative pressure and three-dimensional velocity, while fixed weighted loss terms enforce continuity and momentum balance, no-slip walls, pulsatile inflow, an outlet pressure gauge, and cardiac-cycle periodicity. Predicted fields are postprocessed into WSS, TAWSS, OSI, and RRT descriptors for downstream rupture-risk prediction.

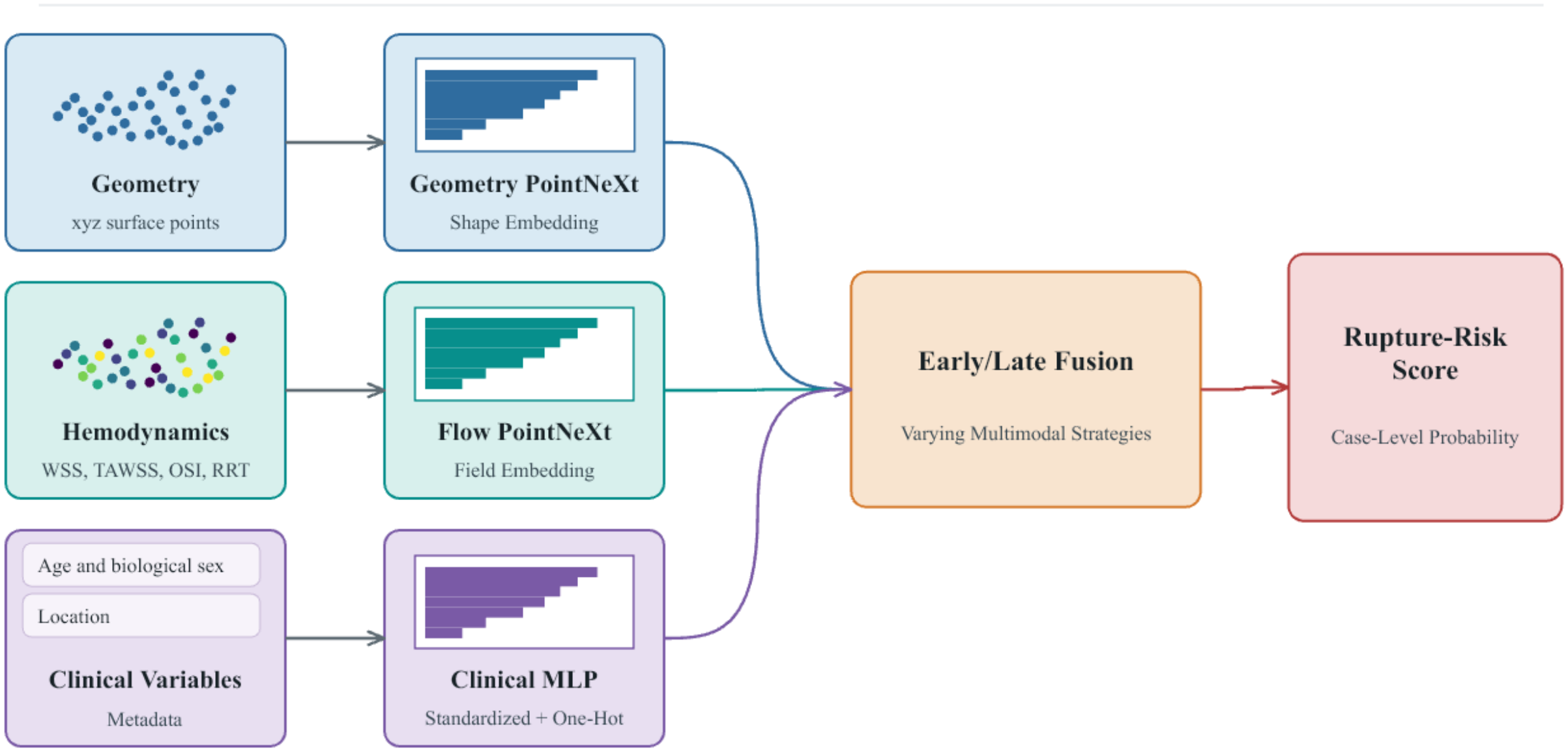

**Supplementary Figure S3. PointNeXt-based multimodal framework for rupture-risk prediction**. Geometry and hemodynamic inputs are encoded by separate PointNeXt branches, while clinical variables are processed by a multilayer perceptron. The hemodynamic branch uses WSS-, TAWSS-, OSI-, and RRT-based channels. Early fusion combines modality-specific embeddings in one prediction head, whereas late fusion combines independently trained Geometry + Clinical and Flow + Geometry probabilities to produce a case-level rupture-risk score.

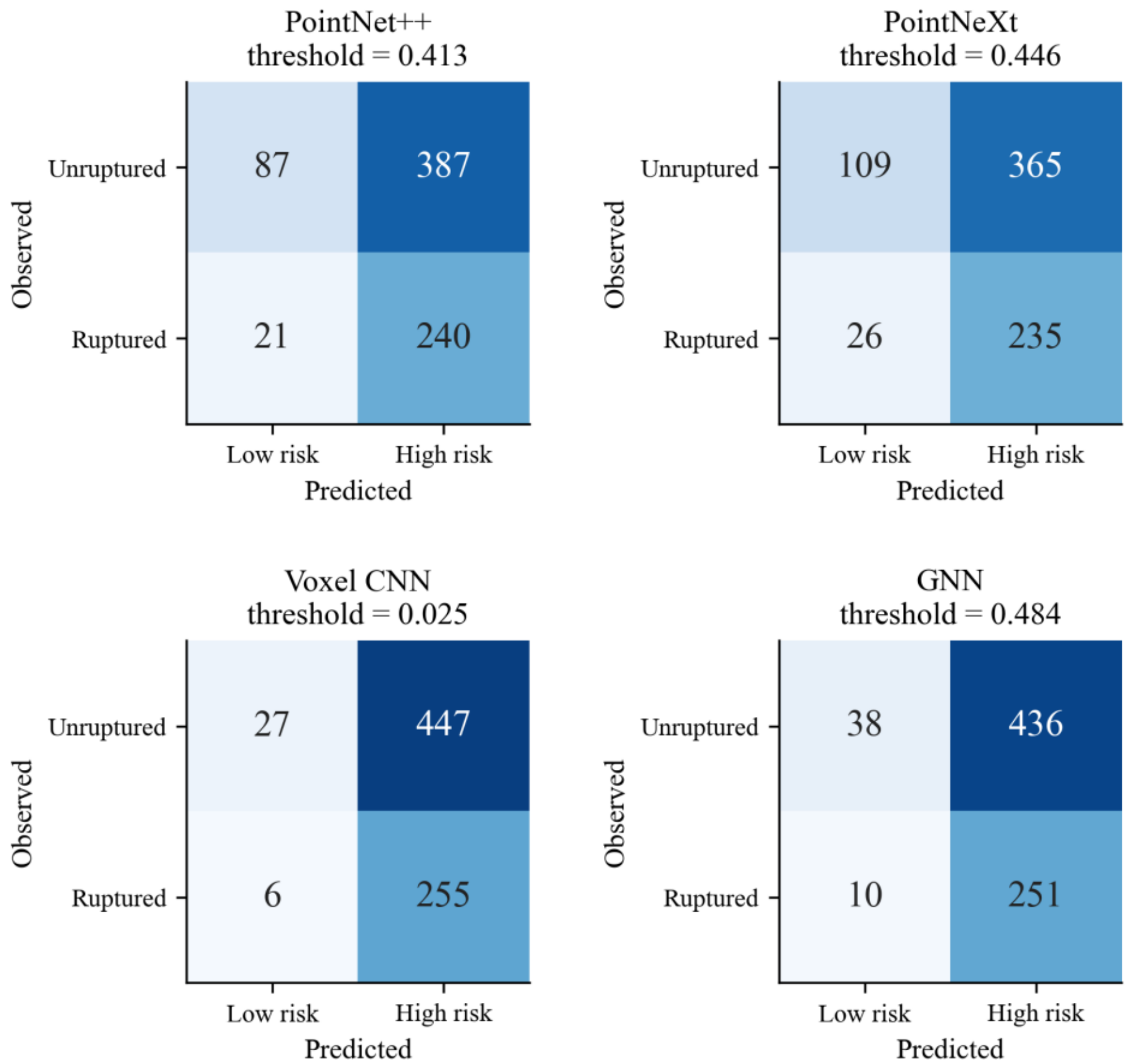

**Supplementary Figure S4. Geometry-only pooled confusion matrices**. Matrices are shown at model-specific F1-optimized thresholds for PointNet++, PointNeXt, voxel CNN, and GNN. Rows denote observed unruptured and ruptured labels, and columns denote predicted labels. Diagonal cells represent correct predictions; off-diagonal cells show false-positive and false-negative errors and should be interpreted alongside the threshold-independent AUROC and AUPRC results.

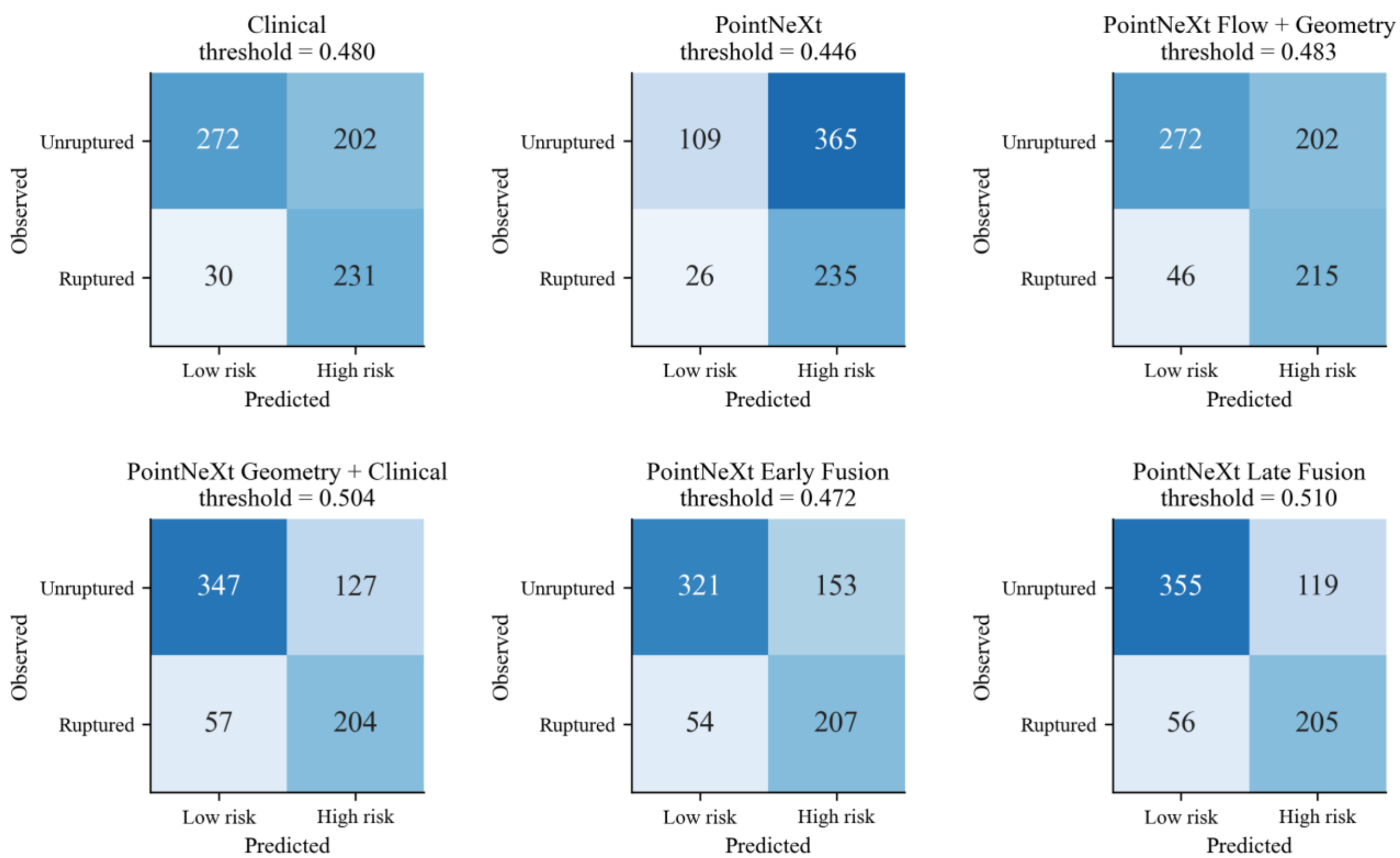

**Supplementary Figure S5. Multimodal pooled confusion matrices.** Multimodal pooled confusion matrices. Matrices are shown at model-specific F1-optimized thresholds for clinical-only, PointNeXt geometry, Flow + Geometry, Geometry + Clinical, early fusion, and fixed 70/30 late fusion. Rows denote observed unruptured and ruptured labels, and columns denote predicted labels. The matrices summarize threshold-dependent error patterns and complement the pooled AUROC and AUPRC comparisons.